\documentclass[11pt]{article}
\usepackage{moriond,epsfig}

\def\Journal#1#2#3#4{{#1} {\bf #2}, #3 (#4)}

\def\PLB{{\em Phys. Lett.}  B}
\def\PRL{\em Phys. Rev. Lett.}

\def\AJ{\em Astrophys. J.} 

\begin{document}
\vspace*{4cm}
\title{THE ALPHA MAGNETIC SPECTROMETER ON THE INTERNATIONAL SPACE STATION}

\author{C. PALOMARES}

\address{Divisi\'on de Astrof\'{\i}sica de Part\'{\i}culas. CIEMAT. 
Avda Complutense 22, 28040 Madrid, Spain}

\maketitle\abstracts{The Alpha Magnetic Spectrometer (AMS) is a 
particle physics detector designed to operate on the 
International Space Station (ISS). 
The aim of AMS is the direct detection of charged particles in the rigidity 
range from 0.5 GV to few TV to perform high statistics studies of cosmic rays 
in space and search for antimatter and dark matter. This will be possible 
because of the large geometrical acceptance ($\sim$ 0.45 m$^{2}$sr), 
a very accurate energy determination and a very precise
particle identification through redundant measurements of its energy, 
velocity and electric charge. Most of the detector components were tested  
with a prototype of the final detector during a
precursor flight on board of the Space Shuttle in 1998.
AMS is scheduled to be placed on the ISS 
at the beginning of 2008 for a 3 year exposure.}

\section{Introduction}

AMS is a multi-purpose experiment intended to perform measurements of charged 
cosmic rays in space (at about 450 km height) during a 3 year mission. 
AMS will search for primary antimatter and dark matter 
contents of the Universe with a sensitivity never reached before.
In addition, the high statistics study of the cosmic ray elements and 
isotopes and their energy flux in a wide energy range
will result in a greatly improved understanding of the cosmic ray 
propagation in our Galaxy.

\begin{figure}
\begin{center}
\includegraphics[height=6.5cm,width=6.5cm]{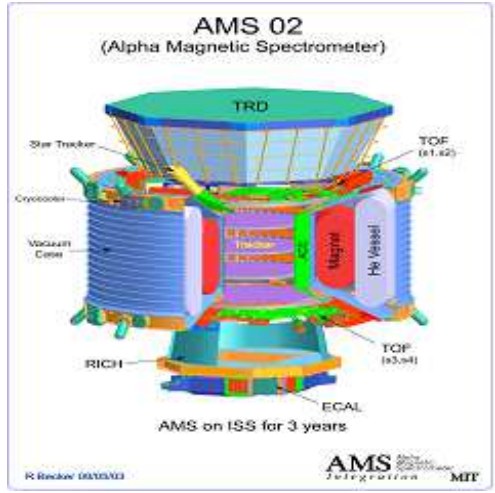}
\end{center}
\caption{AMS-02 detector}
\label{ams_layout}
\end{figure}

\section{AMS detector}
AMS is a magnetic spectrometer equipped with several subdetectors devoted to 
the identification of the particle and the measurement of its energy (see Fig.
\ref{ams_layout}).
The detector strategy is based on redundant measurements which ensure 
a proper background rejection for rare signal searches.
In addition, the design has to meet 
the specific constraints imposed by the launch and the operational 
enviroment conditions.

The main components of the detector are:

\begin{itemize}

\item A cryogenic superconducting magnet that consists of an arrangement of 2 
main dipole coils and 12 racetrack coils designed to give the maximum field in
the perpendicular direction, while minimising the stray field outside the 
magnet. The magnetic flux density at the geometric centre of the system is 
0.86 T.

\item A silicon tracker detector (STD) made of 8 double-sided silicon sensor
layers, 6 of them
contained inside the magnet, will measure the charged particle bending.

\item A time of flight system (ToF) consisting of 2 double planes of 
scintillator counters which is able to reach a precision in the time 
of around 120 ps

\item A transition radiation detector (TRD) which detects the transition 
radiation light 
produced by ultra-relativistic particles in a set of polypropylene fiber 
radiators by means of 5248 straw tubes operated at high voltage with a 
mixture of Xe and CO$_{2}$.

\item A ring imaging \v{C}erenkov counter (RICH) will measure the velocity of 
relativistic particles from its \v{C}erenkov cone opening angle. The 
\v{C}erenkov 
radiator consists of a set of aerogel and NaF tiles, while the detection plane
is instrumented with 680 R7600-M16 multianode Hamamatsu photo-multipliers.

\item An electromagnetic calorimeter (ECAL) that consists of layers of 
lead foils with glued scintillating fibers resulting into a total radiation
depth of 17 $\rm{X}_{o}$ for shower development.
\end{itemize}

The STD mesures the particle rigidity (p/Z) 
in the range from 0.5 GV to 2 TV with a
resolution of 1.5$\%$ at 10 GV. The ECAL takes 
charge of the electromagnetic particle identification, determining the 
electron energy with a resolution of 3$\%$ at 100 GeV.

The particle identification is done through its electric charge and its 
mass. The charge is determined by a combined measurement of the 
deposited energy in the ToF and STD planes and by the \v{C}erenkov light 
detected in the RICH from Z=1 to Z$\leq 26$ with very small charge confusion.
The RICH will also provide a very precise velocity measurement,
$\sigma \approx 0.1\%$ for protons with $\beta > 0.95$, which allows a 
resolution in the mass determination of 2$\%$. 
Below the RICH threshold the velocity is 
measured by the ToF with a resolution of 3.5$\%$.  
 
In addition, the different response of hadronic and electromagnetic particles 
in the interaction with the TRD and ECAL provides a hadron-electron 
separation: The TRD differentiates protons from electrons with a rejection
factor of $10^{2}-10^{3}$ in the range from 1.5 to 300 GeV while the ECAL 
provides a rejection factor of $10^{4}$ for electrons with energy smaller
than 1 TeV.

\section{Galactic Cosmic Rays}

The high energy cosmic ray nuclei are accelerated particles that travel 
through the interstellar medium (ISM), where they are scattered, 
reaccelerated and lose energy before reaching the Earth. 
They can also produce secondary nuclei by fragmentation of heavier ones
(spallation), decay and even escape from the Galaxy. 
The propagation models must provide a reliable description of the CR 
propagation through the ISM taking into account all these fenomena and
reproduce the observed CR fluxes in the heliosphere.
A very precise knowledge of the elemental and isotopic spectra of CR in a 
wide energy range is essential to understand their origin, the matter content
and the 
magneto-hydrodynamical properties of our Galaxy.

AMS will be able to cover a set of critical measurements with very high 
precision, extending significantly their energy range.

\subsection{H and He spectrum}

Hydrogen and helium constitute about 99$\%$ of the hadronic CR,
their energy spectrum provides information about the primary 
acceleration mechanism.
Differences in the origin and acceleration among species could 
be obtained by comparing both high statistics spectra.
Precise 
measurements of H and He fluxes are used to determine the expected fluxes of 
antiprotons and positrons, to compute the diffuse gamma-ray background
spectrum and to define the expected fluxes of atmospheric neutrinos.
The most precise measurements come from magnetic spectrometres with
uncertainties of about 5-10$\%$ up to rigidities of 100-200 GV~\cite{1}. 
AMS should be included
in this set of experiments but improving the resolution and increasing the 
dynamic range up to 2-3 TV (see Fig.~\ref{capabilities}). 
Other experiments, as calorimeters and nuclear emulsions~\cite{2}, have 
measured 
the H spectrum up to 10$^{15}$eV/n but with poor accuracy, i.e., 25-50$\%$.

\subsection{Secondary CR spectrum}

AMS will measure the energy spectrum of primary CR up to Fe, but 
it will keep a larger sensitivity for C, N and O, 
the most abundant elements after H and He. A very interesting measurement
is the ratio of secondary particles produced by spallation to
the corresponding primary one, like for example B/C. It gives information 
about the 
amount of matter traversed by the CR and, therefore, about the 
confinement volume.
Boron to carbon ratio has been obtained by several experiments in different
energy ranges from 10$^{-1}$ to 10$^{2}$ GeV/n~\cite{3}~\cite{4}~\cite{5}. 
Until now the smallest 
uncertainties in the ratio come from HEAO-3 experiment, 
$\sim$ 5$\%$ for $0.6<E<35$ GeV/n. 
AMS will measure the B/C ratio 
from 0.2 to 1000 GeV/n with a resolution better than 5$\%$ in the whole range. 

\subsection{Unstable Isotopes}

\begin{figure}
\begin{center}
\includegraphics[height=5.5cm,width=6.cm]{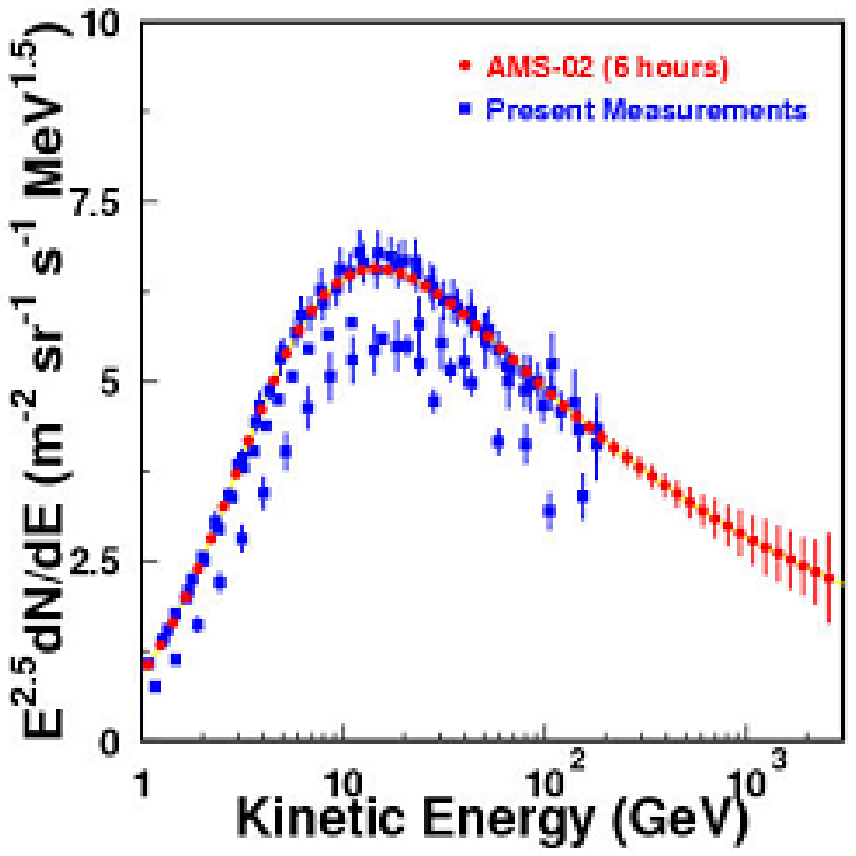}
\includegraphics[height=5.5cm,width=6.5cm]{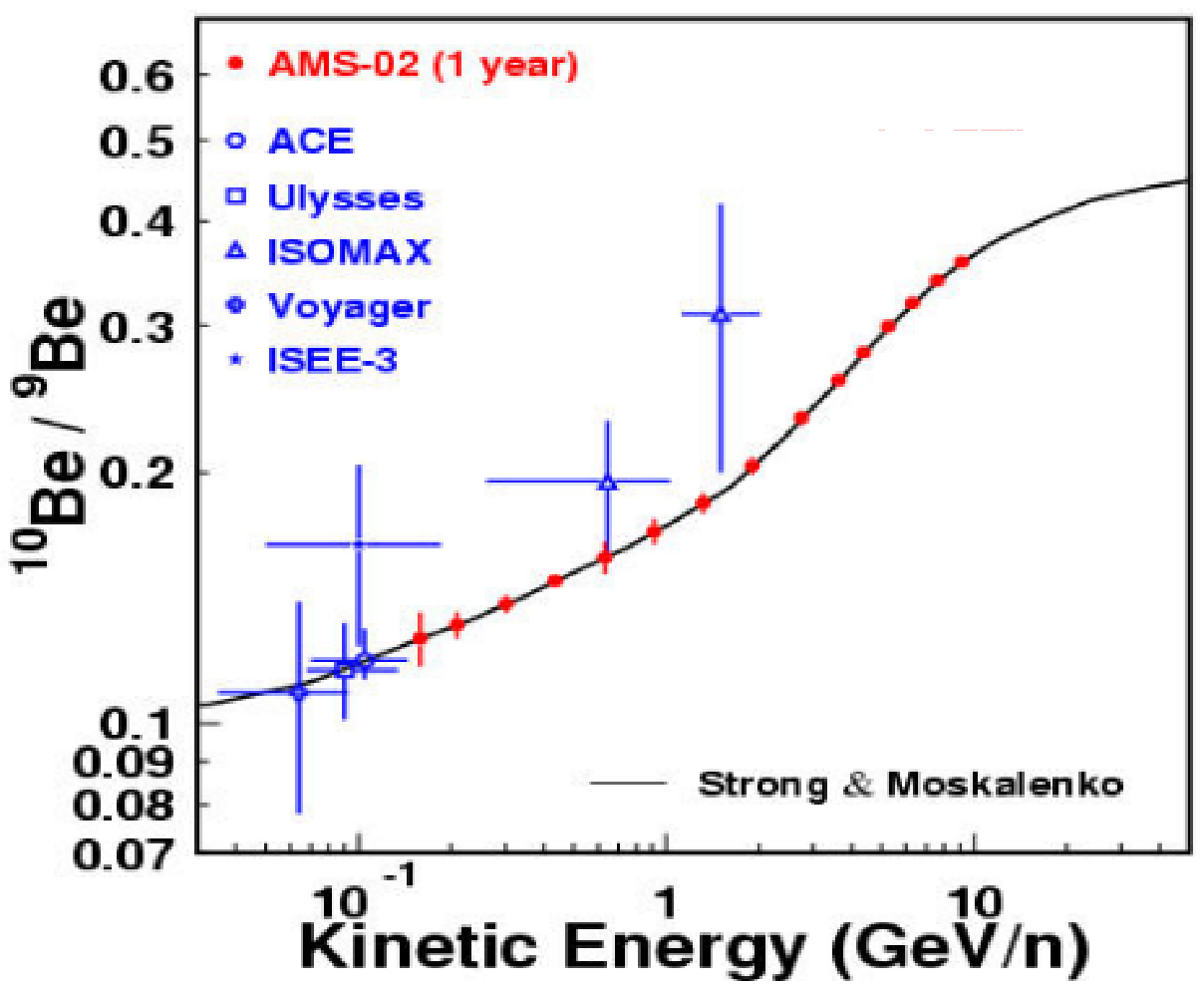}
\end{center}
\caption{AMS-02 expected perfomaces: H spectrum (left) and the ratio 
$^{10}$Be/$^{9}$Be (right)}
\label{capabilities}
\end{figure}

Some of the species created by spallation are radioactive.
$^{10}$Be is particularly
interesting because its half time is of the same
order than the confinement time of the CR in the Galaxy.
The relative abundances of the isotopes of Be can tell us whether or not all
the $^{10}$Be have decayed, and consequently provide an estimate of the 
mean age of the CR observed at the Earth.
Present measurements of the ratio $^{10}$Be/$^{9}$Be
have been performed using 
space-borne spectrometers for energies $\leq 100$ MeV/n
and ISOMAX 
balloon-borne magnetic spectrometer~\cite{7} for energies in the range 
0.26--2 GeV/n, as it is shown in Fig.~\ref{capabilities}.
AMS will be able to separate $^{10}$Be from $^{9}$Be for 
$0.15 ~{\rm GeV/n} < E< 10 ~{\rm GeV/n}$,
the sensitivity after 1 year of data taking is shown in Fig.~\ref{capabilities}.

\subsection{Antiparticles}

Antiparticles, like antiprotons and positrons, are secondary CR 
created by interaction of primary CR with the ISM. They constitute 
the best signature for new physics because of the low background. Distortions
in the antiparticle spectrum, related to its secondary origin, could arise 
from primary source contributions such as neutralino annhilation in the 
galactic halo, hence a very good understanding of the expected fluxes becomes
necessary. 
The detection of antiparticles is reserved to magnetic spectrometers that
have the capabilities for charge sign identification.
The current measurements of the antiproton spectrum are dominated by 
the balloon-borne spectrometer BESS~\cite{8}. These measurements together with 
the ones obtained by AMS-01~\cite{9} and CAPRICE~\cite{10} 
agree with the theoretical predictions for a pure secondary origin.
For positrons, the experimental results are not conclusive. In spite of 
the most recent measurements~\cite{11}, generally compatible with a secondary 
origin, there is some
indication of a structure around 7 GeV in the energy dependence of the 
positron fraction as measured by HEAT.
Only new measurements over a larger energy range and more statistics will
be able to better constrain propagation models and give a clear signal 
physics. After 3 years AMS will collect around 10$^{6}$ antiprotons and
will identify and measure positrons up to 400 GeV. This flux
will provide sensitivity to new physics in several scenarios.

\end{document}